\documentclass[pre,groupedaddress,showkeys,showpacs,twocolumn]{revtex4}
\usepackage{amsmath}
\usepackage{graphics}
\usepackage{graphicx}
\usepackage{amsfonts}
\usepackage{amssymb}
\usepackage{dcolumn}
\usepackage{bm}
\usepackage{natbib}
\begin{document}
\title{Friction of viscoelastic elastomers with rough surfaces under torsional
	contact conditions}
\author{Miguel Trejo}
\author{Christian Fr\'etigny}
\author{Antoine Chateauminois}
\email[]{antoine.chateauminois@espci.fr}
\affiliation{Soft Matter Science and Engineering Laboratory (SIMM), UMR CNRS
7615,
Ecole Sup\'erieure de Physique et Chimie Industrielles (ESPCI), Universit\'e Pierre et Marie Curie, Paris (UPMC), France}
\begin{abstract}
Frictional properties of contacts between a smooth viscoelastic rubber and rigid surfaces are investigated using a torsional contact configuration where a glass lens is continuously rotated on the rubber surface. From the inversion of the displacement field measured at the surface of the rubber, spatially resolved values of the steady state frictional shear stress are determined within the non homogeneous pressure and velocity fields of the contact. For contacts with a smooth lens, a velocity dependent but pressure independent local shear stress is retrieved from the inversion. On the other hand, the local shear stress is found to depend both on velocity and applied contact pressure when a randomly rough (sand blasted) glass lens is rubbed against the rubber surface. As a result of changes in the density of micro-asperity contacts, the amount of light transmitted by the transparent multi-contact interface is observed to vary locally as a function of both contact pressure and sliding velocity. Under the 
assumption that the intensity of light transmitted by the rough interface is proportional to the proportion of area into contact, it is found that the local frictional stress can be expressed experimentally as the product of a purely velocity dependent term, $k(v)$, by a term representing the pressure and velocity dependence of the actual contact area, $A/A_0$. A comparison between $k(v)$ and the frictional shear stress of smooth contacts suggests that nanometer scale dissipative processes occurring at the interface predominate over viscoelastic dissipation at micro-asperity scale.
\end{abstract}
\maketitle
\section{Introduction}
\label{sec:Introduction}
Rubber friction is a topic of huge practical importance in many applications, such as tires, rubber seals, conveyor belts, and syringes, to mention only a few. However, there is an incomplete understanding of the parameters that control the frictional behavior of rubber surfaces. Since the seminal experimental work by Grosh~\cite{grosch1963a}, rubber friction is usually assumed to involve two dissipative components. The first one, often denoted as the adhesive component, corresponds to thermally and stress activated pinning/depinning mechanisms between rubber molecules and the contacting surface. This idea form the basis of the Schallamach model~\cite{schallamach1963} which was subsequently extended by Chernyak and Leonov~\cite{Chernyack1986}. In a later study, Vorvokalos and Chaudhury~\cite{vorvolakos2003effects} also showed that these models can consistently be used to describe the dependence of friction of poly(dimethylsiloxane) (PDMS) elastomers on molecular parameters such as molecular weight. The 
second dissipative component involved in rubber friction is assumed to correspond to viscoelastic losses associated with the contact deformation of the soft rubber. In the case of a hard, rough surface sliding on a viscoelastic rubber, viscoelastic losses at microasperity scale occurs at characteristic frequency of the order of $v/d$ where $v$ is the sliding velocity and $d$ is a characteristic size of asperity contacts. This so-called hysteretic component to friction was first evidenced by Greenwood and Tabor~\cite{greenwood1958} in a series of experiments, in which hard spheres and cones were sliding or rolling on well-lubricated rubber surfaces. The work by Grosh~\cite{grosch1963a} extended these investigations to the more complex situation of rubber sliding on microscopically rough surfaces. A maximum in friction was found to occur at a sliding velocity related to the frequency with which the asperities of the rough surface deform the rubber surface. This maximum was absent on a smooth track, thus 
reflecting the deformation losses induced by 
the passage of the asperities over the rubber surface. These frictional mechanisms involving viscoelastic losses at microasperity scale have motivated the development of several theoretical models starting from Fourier transform analysis applied to periodic surfaces~\cite{Schapery1978,golden1980} to the more complex model developed by Persson for rubber friction on randomly rough surfaces~\cite{Perrson2006,persson2001}. Using a spectral description of the topography of the rough surfaces, Persson's theory predicts how the component of friction force associated with hysteretic losses varies with velocity and contact pressure from an estimate of the actual contact area. Some experimental results tend to support this theory~\cite{lorenz2011} but a detailed examination of the effects of surface topography on rubber friction remains very challenging in the case of randomly rough surfaces where adhesive and hysteretic components are strongly intricate.\\
In a previous work~\cite{nguyen2013}, we have investigated the friction of a PDMS rubber with model rough surfaces consisting of silica lenses covered with various densities of spherical colloidal nano-particles. From an examination of the pressure dependence of the frictional shear stress, we showed that the actual contact area was close to saturation in the whole range of applied contact load and sliding velocity. These model surfaces thus allowed quantifying the contributions of interface dissipation and hysteretic losses to friction without the complications arising from the pressure and velocity dependence of the actual contact area. In addition, the use of a monodisperse distribution of colloidal particles allowed to control both the characteristic frequency associated with deformation at asperity scale and the volume of the viscoelastic substrate that is affected by this deformation. Within this framework, we were able to determine experimentally the hysteretic component of friction which compares 
well with theoretical calculations. In this study, we consider the more realistic situation of a viscoelastic rubber sliding against a randomly rough rigid surface where the proportion of area into contact is expected to depend on both the applied pressure and the sliding velocity. Experiments are carried out using a torsional contact configuration which -as explained below- allows investigating frictional energy dissipation at the interface without the complications arising from bulk viscoelastic losses at the scale of the macroscopic contact. In addition, the inversion of the measured displacement field at the surface of the rubber provides local values of the frictional shear stress within the non homogeneous pressure and sliding velocity fields of the contact. Local changes in the density of asperity micro-contacts are evidenced from a measurement of the amount of light transmitted through the transparent rough contact.\\
In a first part of this paper, we consider the case of a smooth contact where friction is likely to arise only from molecular scale dissipation at the intimate contact formed between the surfaces. In a second part, we examine the pressure and velocity dependence of the frictional shear stress within rough contacts where asperity scale viscoelastic losses are likely to come into play. We show that the measured shear stress can be expressed as the product of a velocity dependent term by a velocity and pressure dependent term which describes the changes in the actual contact area as a function of nominal contact pressure and sliding velocity. From a comparison between the smooth and rough contacts, we discuss in a last part the contributions of interface dissipation and hysteretic losses to friction.
\section{Experimental details}
\label{sec:experimental_details}
\subsection{Materials and sample preparation}
\label{subsec:materials}
\begin{figure}[ht!]
	\centering
	\includegraphics[width=\columnwidth]{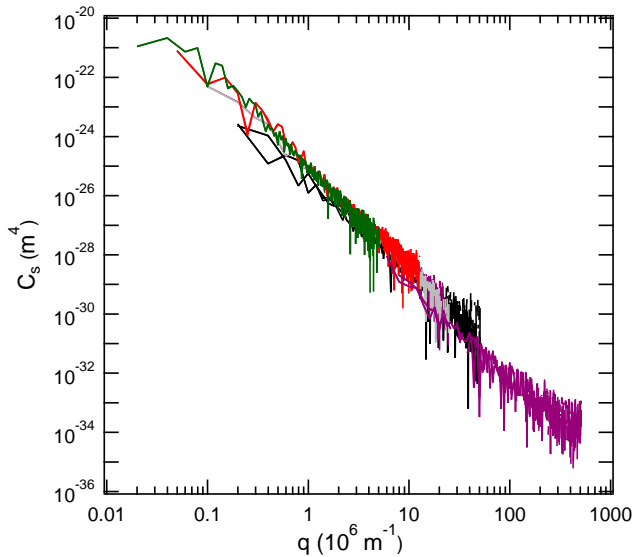}
	\caption{(Color online) Roughness Power Spectrum Density (PSD) of the sand blasted
		glass lens as measured using AFM. Colours denote different image sizes, from left to right:
		$50~\times~50~\mu\text{m}^2$ (green), $20~\times~20~\mu\text{m}^2$ (red),
		$10~\times~10~\mu\text{m}^2$ (gray), $5~\times~5~\mu\text{m}^2$ (black),
		$0.5~\times~0.5~\mu\text{m}^2$ (purple).}
	\label{fig:PSD_rough_lens}
\end{figure}

As a substrate, we use an epoxy based rubber obtained by crosslinking diglycidil ether of bisphenol A (DER 332, $M_w=340~g~mol^{-1}$, Dow Corning) with a polyether-diamine crosslinker (Jeffamine\textsuperscript{\textregistered}$\:$ED2003, $M_w=2003~g~mol^{-1}$, Hunstman Chemical). As detailed in the appendix, this rubber exhibits a significant change (about one order in magnitude) in the loss modulus in the characteristic frequency range ($\approx~0.1-10^3~Hz$) involved in surface deformation at microasperity scale. In order to elaborate the specimens, each of the reactive parts is first separately stirred in a silicone bath at $70\,^{\circ}\mathrm{C}$ during about 30 min. Then, epoxy is mixed with the stoichiometric amount of diamine determined with the epoxy equivalent weight and amine hydrogen equivalent weight given by the supplier (Jeffamine\textsuperscript{\textregistered} Data Sheets). The reactive mixture is stirred and subsequently degassed about 40 min at $50\,^{\circ}\mathrm{C}$ in a vacuum 
chamber. Then, the mixture is poured into a parallelepiped shaped PDMS mold (size: $4.5\ \text{cm}\times4.5\ \text{cm}\times1.5\ \text{cm}$) and cured at $120\,^{\circ}\mathrm{C}$ for 20 h. In order to monitor contact induced surface displacements, a square network of small cylindrical holes (diameter $10\ \mu\text{m}$, depth $2\ \mu\text{m}$ and center to center spacing $70\ \mu\text{m}$) is stamped on the PDMS surface. Once imaged in transmission with a white light, the pattern appears as a network of dark points. This surface marking is simply achieved by patterning the bottom part of the PDMS mold by a network of cylindrical posts using conventional soft lithography techniques. After curing, the glass transition temperature of the epoxy rubber is $-42^{\circ}\mathrm{C}$, as determined by Differential Scanning Calorimetry (DSC) at a scan rate of $10\,^{\circ}\mathrm{C}\:\text{min}^{-1}$.\\
During friction experiments, the rubber specimen is contacting a plano-convex BK7 glass lens (Melles Griott, France) with a radius of curvature of 14.8~mm. After cleaning, the r.m.s. roughness of the lens is less than $2~\text{nm}$, as measured by AFM using $1~\times~1~\mu m^2$ pictures. One of the lenses is rendered microscopically rough using sand blasting (average grain size of 60~$\mu$m). The topography of the surface has been characterized by AFM measurements using image sizes ranging from $50~\times~50~\mu m^2$ to $500~\times~500~nm^2$. Fig.~\ref{fig:PSD_rough_lens} depicts the results in the form of a roughness Power Spectrum Density (PSD) $C_s(q)$. This PSD decays according to a power law, from $50\ \mu\text{m}$ down to the nanometer scale. Accordingly, the surface roughness can be defined as self-affine fractal ($C_s(q)\varpropto q^{-2(H+1)}$) with a Hurst exponent $H=0.58$ and a fractal dimension $D_{f}=3-H=2.42$. The r.m.s roughness of the sand blasted surface is measured as $1.69~\pm~0.19~\mu\
text{m}$ using $50~\times~50\:\mu\text{m}^2$ images.
\subsection{Friction setup and contact imaging}
\label{subsec:friction_setup}
\begin{figure}[t!]
	\includegraphics[width=\columnwidth]{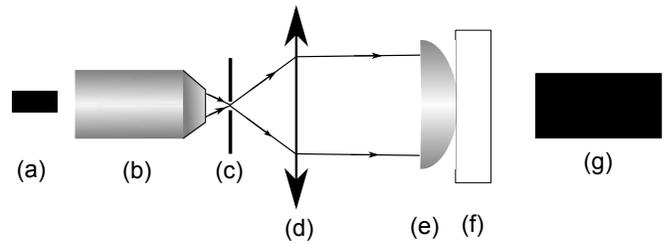}
	\caption{Schematic description of the parallel light system used to illuminate the contact region. (a) optical fiber light; (b) microscope objective; (c) pinhole; (d) convex lens; (e) contact lens; (f) rubber specimen; (g) CMOS camera.}
	\label{fig:setup}
\end{figure}

\begin{figure}[t!]
	\includegraphics[width=\columnwidth]{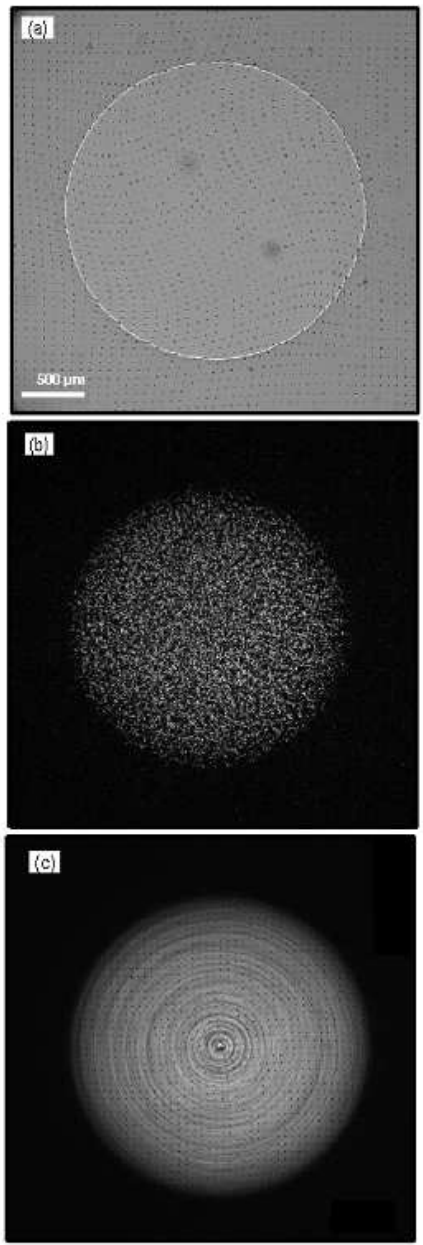}
	\caption{Contact pictures taken during steady state friction. (a) smooth contact; (b) rough contact; (c) rough contact after image averaging. Averaging allows suppressing most of the light intensity fluctuations due to the rotating rough surface. The dot lattice on the rubber surface then becomes apparent.}
	\label{fig:contact_pictures}
\end{figure}

Contact torsion experiments are carried out using a custom made device which is fully described in reference~\cite{chateauminois2010}. The experiments consist in rotating continuously a glass lens about an axis perpendicular to the surface of the rubber substrate and passing through the apex of the lens. Normal contact is achieved under imposed indentation depth condition (between 60 and 320~$\mu\text{m}$) by means of a linear displacement stage. The resulting contact radius lies in the range 0.3-2.2~mm. Specimen size ($4.5\ \text{cm}\times4.5\ \text{cm}\times1.5\ \text{cm}$) ensures that the ratio of the substrate thickness to the contact radius is greater than ten, i.e. that semi-infinite contact conditions are achieved during torsion experiments~\cite{Gacoin2006a}. Separate indentation experiments using the same device equipped with a load cell allowed to determine the relationship between indentation depth and normal load (the load cell has to be removed during torsional contact experiments for imaging 
purposes). During friction experiments, the glass lens is rotated at imposed angular velocity between 0.01 and 10~deg~s~$^{-1}$ using a motorized rotation stage. Prior to use, the lenses are successively cleaned with acetone and ethanol in an ultrasonic bath during about 5 min. Epoxy based specimens are thoroughly washed with 2-isopropanol and subsequently dried under vacuum.\\ During torsion, images of the contact zone are continuously recorded through the transparent rubber substrate using a zoom lens and a CMOS camera. The system is configured to a frame size of $1024~\times~1024$ pixels with 8 bits resolution. Images are acquired at a frequency ranging from 0.01 to 30 Hz. The contact zone is illuminated using a parallel light system located behind the glass lens, as schematically described in Fig. \ref{fig:setup}. In the case of the smooth contact interface, subpixel detection of individual markers on the epoxy surface is carried out directly from single images taken during steady state friction (as 
that shown in Fig.~\ref{fig:contact_pictures}a) using a particle tracking method. Each contact picture provides a displacement field with about 6,000 data points with a spatial resolution corresponding to distance between markers (i.e. $70\ \mu\text{m}$). In the case of rough interfaces, the contact appears as bright spots against a darker background as a result of light scattering by the roughened surface (Fig.~\ref{fig:contact_pictures}b). It is therefore no longer possible to detect the markers on the rubber surface on a single image. However, an averaging procedure allows revealing the location of the markers under steady state friction. As shown in Fig.~\ref{fig:contact_pictures}c, averaging several images taken during steady state friction suppresses nearly all the light intensity fluctuations induced by surface roughness thus allowing to reveal the location of the markers which are fixed with respect to the camera.
\section{Friction of smooth contacts}
\label{sec:smooth}
\begin{figure}[t!]
	\includegraphics[width=0.9\columnwidth]{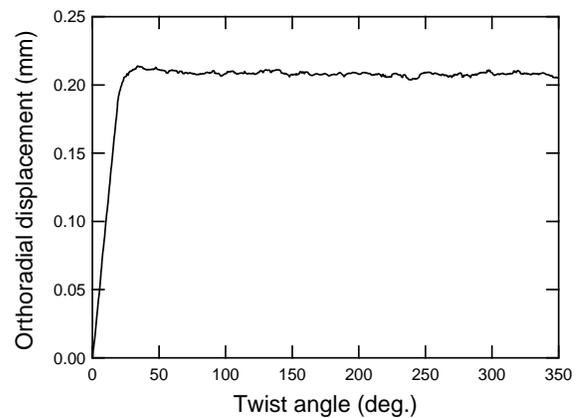}
	\caption{Azimuthal displacement of an individual marker on the surface of the rubber substrate as a function of the applied twist angle (angular displacement
		rate: 3~deg~s$^{-1}$). The marker is located at a radial coordinate $r=0.63\:\text{mm}$, the contact radius is $a=1.12\:\text{mm}$ (smooth lens).}
	\label{fig:utheta_theta}
\end{figure}

\begin{figure}[t!]
	\includegraphics[width=\columnwidth]{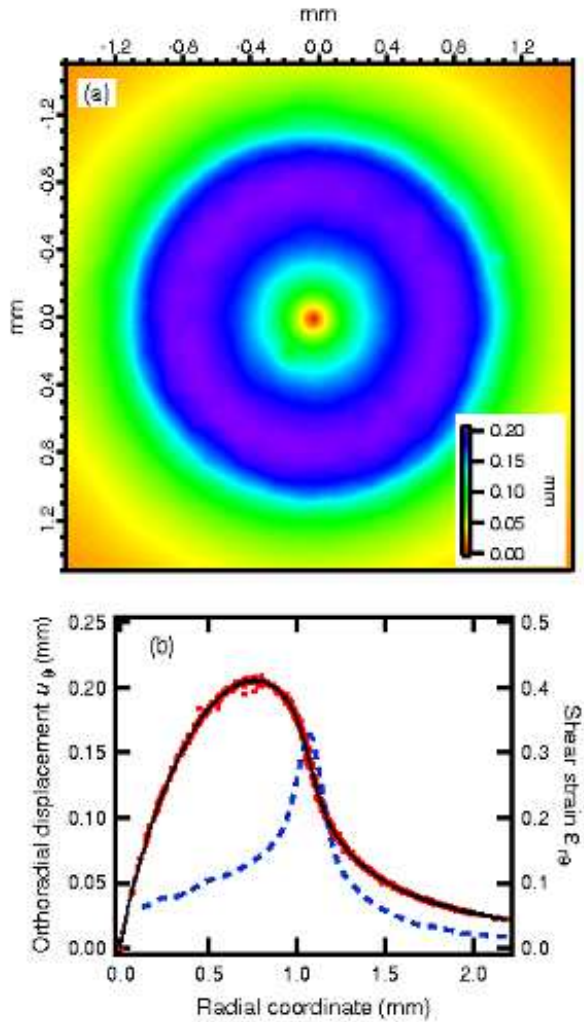}
	\caption{(Color on line) Steady state displacement field within a smooth contact (angular velocity:~$3\:\text{deg s}^{-1}$, indentation depth: 100~$\mu$m).(a) image of the azimuthal displacement field; (b) radial profile of the azimuthal displacement (dots) and of the associated surface shear strain $\epsilon_{r \theta}$ (blue dotted line). The plain black line corresponds to the fit of $u_{\theta}$ which was used to calculate the surface strain.}
	\label{fig:displ_field}
\end{figure}

\begin{figure}[t!]
	\includegraphics[width=\columnwidth]{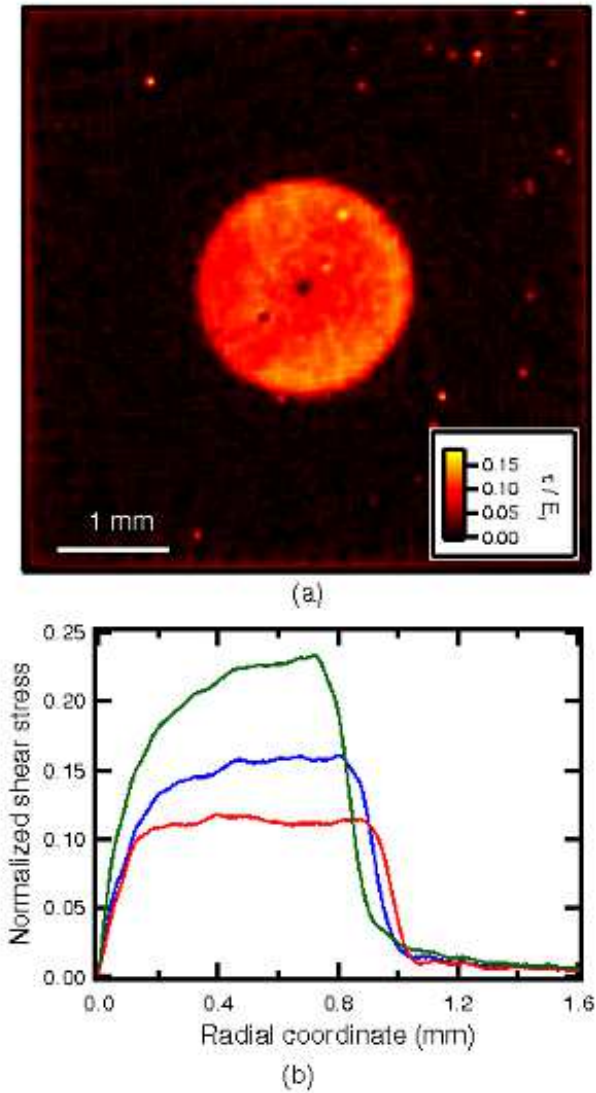}
	\caption{(Color online) Steady state frictional stress within a smooth contact interface (indentation depth: 60~$\mu$m).(a) image of the shear stress distribution (angular velocity: $0.3\:\text{deg s}^{-1}$); (b) radial profiles of the shear stress. Angular velocities from bottom to top: 0.3, 1 and 3~deg~s$^{-1}$. The shear stress is normalized with respect to the relaxed modulus, $E_r$, of the rubber.}
	\label{fig:stress_field}
\end{figure}

\begin{figure}[t!]
	\includegraphics[width=\columnwidth]{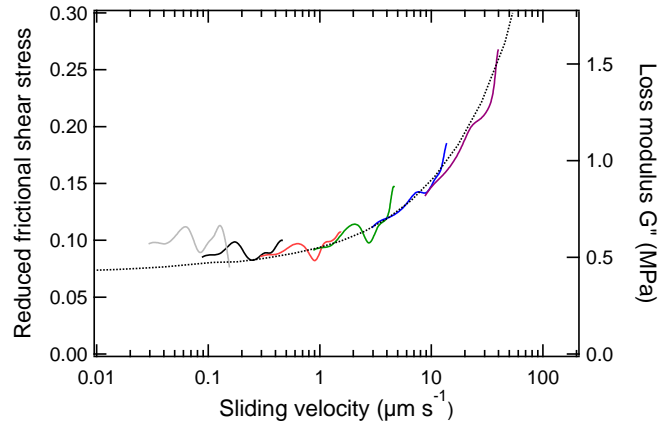}
	\caption{(Color online) Master curve giving the reduced local shear stress as a function of the local sliding velocity within a smooth contact. Imposed velocities from left to right:
		0.01, 0.03, 0.1, 0.3, 1 and 3~$\:\text{deg s}^{-1}$. The dotted line represents the change in the loss component of the shear modulus $G''$ as a function of $\lambda \omega$, where $\omega$ is the frequency and $\lambda$ is a typical length set to 6~nm (the curve has been shifted horizontally in order to allow a comparison with the shape of the reduced shear stress plot).}
	\label{fig:master_smooth}
\end{figure}

When the smooth lens is twisted starting from rest, a stiction stage is first encountered which corresponds to the shear failure of the adhesive contact. This stiction process occurs according to a fracture like process characterized by progressive slip propagation from the periphery to the center of the contact. This phenomenon was discussed in a previous study \cite{chateauminois2010} and it will not be considered further in this paper. In the case of the investigated epoxy rubber, this transient stiction phenomenon occurs for twist angles $\theta_{s}\lesssim 50\ \text{deg}$ for all indentation depths and angular velocities under consideration. Then, a steady state friction state is achieved as indicated by the time independence of markers location on the rubber surface. As an example, Fig.~\ref{fig:utheta_theta} depicts the displacement of an individual marker located within the contact as a function of the applied twist angle. Owing to the symmetry the contact, this displacement is expressed using its 
cylindrical components with respect to the center of rotation (only the azimuthal displacement component $u_{\theta}$ is reported in the figure as the radial displacement component $u_{r}$ is found to be systematically negligible in all experiments). After an initial increase corresponding to the stiction stage ($\theta \lesssim 40\:\text{deg}$), a steady state is achieved. Here, the time independent location of the marker is indicative of the achievement of a vanishing strain rate within the bulk rubber substrate. This means that no significant relaxation process takes place at the scale of the contact within the considered time window. The bulk substrate can thus be considered as deformed in a relaxed, time-independent, state.\\
Figure~\ref{fig:displ_field} shows a typical displacement field obtained with the smooth contact under such a steady state friction condition. From this measured displacement field, the corresponding contact stress distribution can be retrieved using an appropriate inversion procedure. In a previous study dealing with linear sliding of silicone rubbers~\cite{nguyen2011}, we showed that an inversion method based on a linear elastic contact mechanics approach can be inaccurate due to the occurrence of finite strains at the edge of the contact. A Finite Elements (FE) inversion procedure was thus developed in order to handle the associated geometrical and material non linearities. Here, a calculation of the surface shear strain
$\epsilon_{r \theta} = 1/2 \left( \partial u_{\theta} / \partial r  - u_{\theta}/r \right)$ from the measured azimuthal displacement profiles (bottom part of Fig.~\ref{fig:displ_field}) shows that strain as high as 0.3 are achieved at the vicinity of the contact edge which are also outside the linear range of the epoxy rubber (about 0.1). In order to evaluate the effects of these non linearities on the inversion, a displacement field was inverted using either a linear elastic approach based on Green's tensor or a FE method able to handle the geometrical and material linearities of the problem. The results reported in appendix B show that both approaches give the same result. It therefore turns out that finite strains do not induce any significant error in a linear elastic inversion of torsional displacement which can be justified by some theoretical considerations~\cite{huy2013}. As a result, all the stress fields to be reported in this study have been obtained from the semi-analytical deconvolution of the 
measured displacement fields using the Green's tensor approach fully detailed in reference~\cite{chateauminois2008}.\\
The surface shear stress distribution of the smooth contact interface was systematically determined from the inversion of the measured steady-state azimuthal displacements at various indentation depths and angular velocities. All the shear stress data obtained from the inversion are expressed in a non dimensional form, $\bar\tau(r)=\tau_{\theta z}(r)/E_r$, where $E_r$ is the relaxed elastic modulus of the rubber. As shown in Fig.~\ref{fig:stress_field}a, a nearly constant frictional shear stress is achieved within the contact zone except at the center of the contact where shear stress vanishes for symmetry reasons. Contact pressure being expected to decrease continuously along the contact radial coordinate, it turns out that frictional shear stress is pressure independent, as already reported for smooth glass/PDMS contacts~\cite{chateauminois2010,chateauminois2008}. A close examination of stress profiles obtained at various imposed velocities (Fig.~\ref{fig:stress_field}b) shows a systematic positive 
gradient along  the radial coordinate which should reflect the velocity dependence of the interface shear stress. This assumption was further considered from a plot of the measured local shear stress values as a function of the local sliding velocity $v = \dot{\theta}r$, where $\dot{\theta}$ is the angular velocity and $r$ is the radial coordinate. According to a previous investigation~\cite{chateauminois2010}, the transition to a vanishing frictional stress in the vicinity of the contact center occurs over a length scale which represents about 10\% of the contact radius and which is essentially dictated by the cut-off frequency of the deconvolution operation. As a result, data points close to the center of the contact ($r/a<0.1$ where $a$ the contact radius) were discarded from the analysis together with data points outside the contact area ($r/a>1$). As shown in Fig.~\ref{fig:master_smooth}, all the selected shear stress values merge on a single master curve when the applied angular velocity is varied. 
Over nearly three orders of magnitude in the sliding velocity, the shear stress is observed to increase continuously by about a factor three. The shear stress being measured in a steady state friction regime where no displacement occurs at the macroscale, it is thus associated with small scale dissipative processes. For such a smooth and intimate contact, friction is usually considered to arise from molecular scale dissipative processes occurring at the sliding interface. As mentioned in the introduction, formation and breakage of adhesive molecular bonds at the contact interface is often invoked as the underlying physical mechanism~\cite{schallamach1963,Chernyack1986}. For rubber sliding on optically smooth glass, Grosh~\cite{grosch1963a} noted that the velocity corresponding to maximum friction and the frequency corresponding to maximum viscoelastic loss form a ratio that is of the order of 7~nm for various materials. This nanometric length scale was assumed by Grosh to represent the molecular scale 
involved in the pinning and depinning process of molecular chains to the glass surface. Here, the available frequency and sliding velocity ranges do not allow to extract a very accurate value of this characteristic length scale. However, it can be seen in Figure~\ref{fig:master_smooth}  that the shape of the $\bar\tau(v)$ plot matches that of the loss component of the shear modulus, $G``$, when the latter is represented as a function of $\lambda \omega$ where $\omega$ is the frequency and $\lambda$ is a characteristic length close to 6~nm. In the following section, we address frictional dissipative processes occurring at larger length scales, i.e. at the scale of micro-asperity contacts within the rough contact interface.
\section{Friction of rough contacts}
\label{sec:rough}
\subsection{Shear stress field}
\begin{figure}[t!]
	\includegraphics[width=\columnwidth]{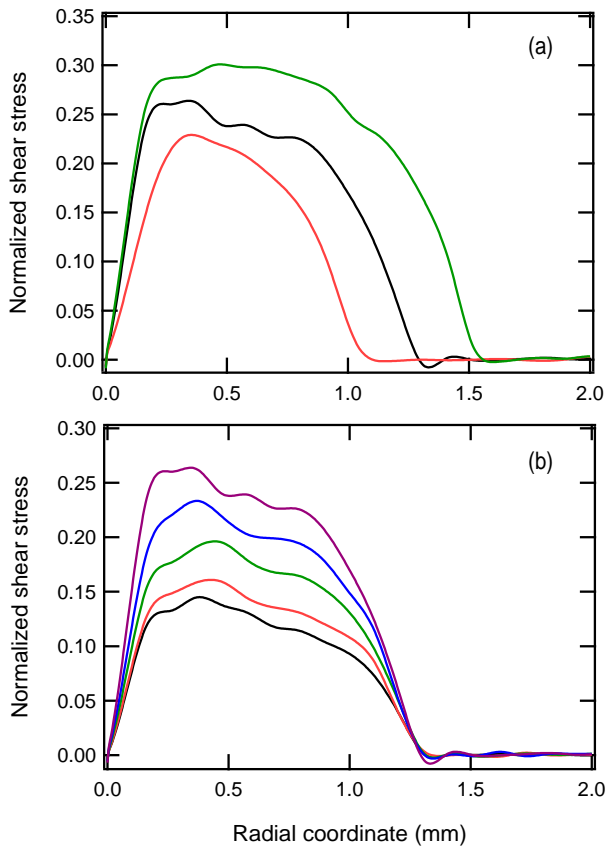}
	\caption{(Color online) Shear stress profiles obtained for various indentation depths and imposed angular velocities. (a) indentation depth, from bottom to top: 60, 100 and 140~$\mu$m; velocity: 1~$\:\text{deg s}^{-1}$. (b) imposed velocity from bottom to top: 0.01, 0.03, 0.1, 0.3 and 1~$\text{deg s}^{-1}$; indentation depth: 100$\:\mu$m.}
	\label{fig:stress_prof_rough}
\end{figure}

In this section, we report on the frictional properties of the contact interface between the smooth viscoelastic elastomer and the sand blasted glass lens. As opposed to smooth contact, a dependence of the local frictional shear stress on contact pressure is now evidenced. As an example, shear stress profiles for various indentation depths are reported in Fig.~\ref{fig:stress_prof_rough}a. The shear stress is clearly decreasing along the radial coordinate, i.e. when the contact pressure decreases. Similarly, increasing applied indentation depths (i.e. contact pressure) result in enhanced shear stress values. Such a pressure dependent frictional stress can be qualitatively accounted by the existence of a multi-contact interface where discrete micro-contacts are distributed within the frictional interface. As the local contact pressure is increased, a higher density of micro-contacts is achieved which in turn results in an enhanced local frictional shear stress. In addition, a velocity dependence of the shear 
stress similar to that observed with smooth contacts is also evidenced (Fig.~\ref{fig:stress_prof_rough}b). Here, the analysis of the local shear stress distribution is complicated by the fact that the local density of microcontacts not only depends on contact pressure but also potentially on the local sliding velocity as a result of viscoelastic effects. In the following section, the changes in the density of micro-contacts as a function of local pressure and velocity is further considered from an examination of fluctuations in the light transmitted by the transparent rough contacts.
\subsection{Optical transmissivity of the multi-contact interface}
\begin{figure}[t!]
	\includegraphics[width=\columnwidth]{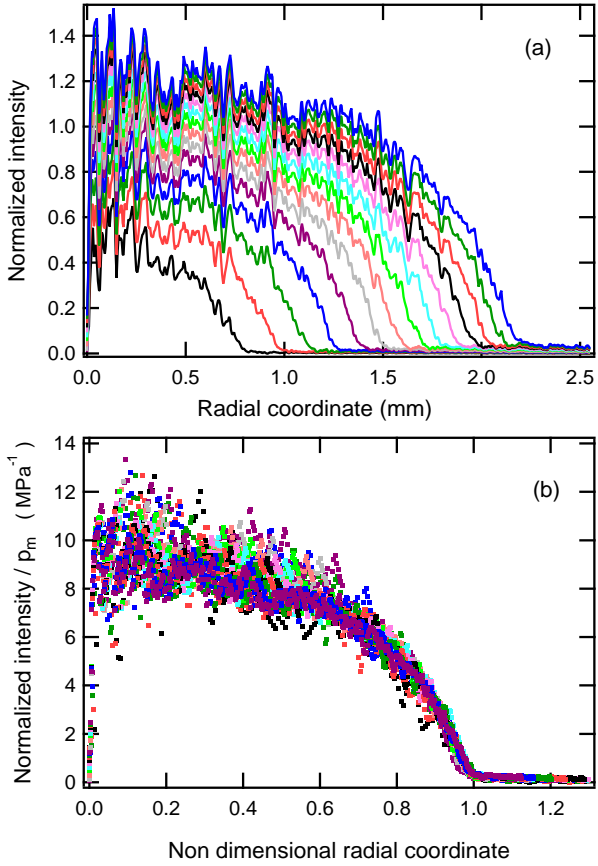}
	\caption{(Color online) Transmitted light intensity during static indentation experiments. (a) normalized intensity profiles obtained at increasing applied indentation depth (from 20 to 300~$\mu$m by 20~$\mu$m steps, from bottom to top). (b) reduced profiles obtained by dividing the normalized intensity and the radial coordinate by the average pressure $p_m=P/\pi a^2$ and the contact radius $a$, respectively ($P$ is applied normal load).}
	\label{fig:intensity_normal}
\end{figure}

\begin{figure}[t!]
	\includegraphics[width=\columnwidth]{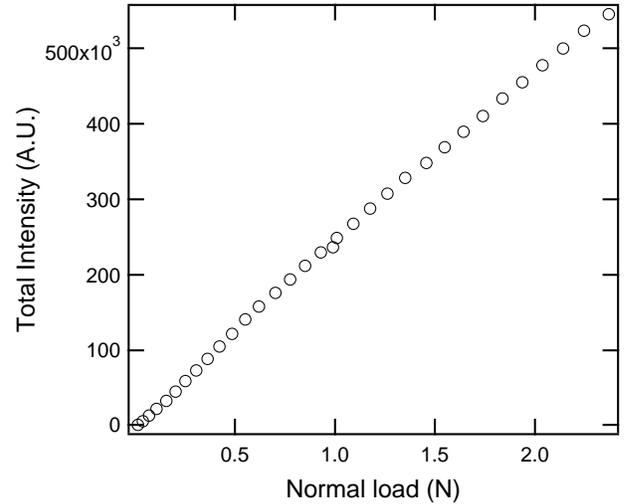}
	\caption{Integrated normalized light intensity transmitted through the rough contact as a function of normal load (static indentation).}
	\label{fig:total_intensity}
\end{figure}

\begin{figure}[t!]
	\includegraphics[width=\columnwidth]{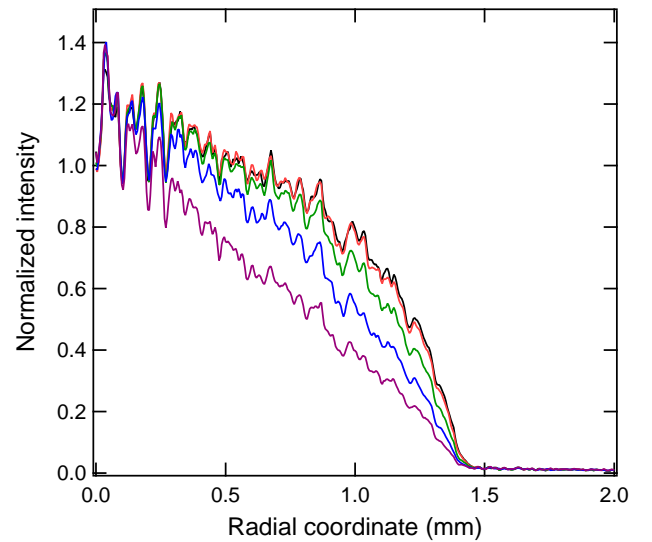}
	\caption{(Color online) Radial profiles of the normalized transmitted light intensity for various imposed angular velocities (indentation depth: 100~$\mu$m). Imposed angular velocities from top to bottom: 0.01~deg~s$^{-1}$ (black), 0.03~deg~s$^{-1}$ (red), 0.1~deg~s$^{-1}$ (green), 0.3~deg~s$^{-1}$ (blue), 1~deg~s$^{-1}$ (purple).}
	\label{fig:intensity_profiles}
\end{figure}

\begin{figure}[t!]
	\includegraphics[width=\columnwidth]{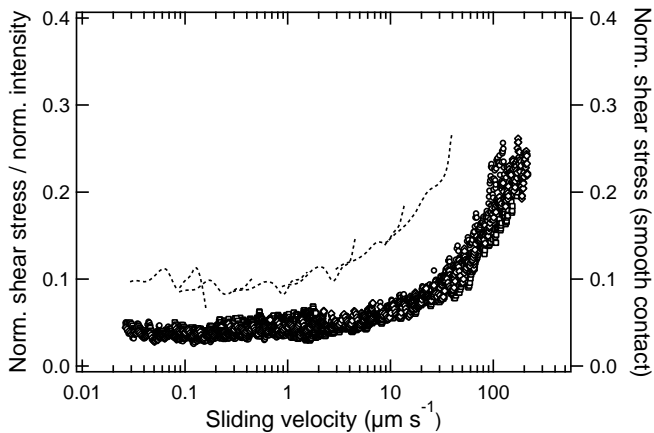}
	\caption{Normalized ratio of frictional shear stress to transmitted light intensity as a function of the sliding velocity. The experiments are carried out using three different indentation depths: ($\circ$) 60~$\mu$m , ($\square$) 100~$\mu$m and ($\diamond$) 140~$\mu$m. The shear stress is normalized with respect to the relaxed modulus of the rubber and the light intensity according to Eqn~(\ref{eq:pix_normalization}). The dotted line correspond to the normalized shear stress measured with the smooth contact (data from Fig.~\ref{fig:master_smooth}).}
	\label{fig:kv}
\end{figure}

Some interesting features of the rough contacts emerge when the changes in the light transmitted locally by the interface are considered. As mentioned above, rough contacts appear as spatially heterogeneous as a result of the scattering nature of the glass surface (cf Fig.~\ref{fig:contact_pictures}b). Because of the difference between the index of refraction of the solids and that of the air, the rough interface transmits light more efficiently when the surfaces are in intimate contact than when they are out of contact. No complete optical model is available to describe these effects but, as a first approach, one can neglect scattering and just consider light transmission in contact and non contact regions of the rough interface. Obviously, light transmission will be more efficient if only one interface is present (contact condition) instead of two (non contact condition). Accordingly, the intensity of transmitted light at a given location within the rough contact should carry informations about the actual 
area of micro-asperity contacts. Such an idea was initially developed by Dietrich and Kilgore~\cite{dieterich1996,dieterich1994} in a study where the actual contact area between rough transparent materials was determined from microscope contact observations. The relevance of this approach to rough contacts interfaces involving polymers was subsequently demonstrated in later studies by Scheibert \textit{et al.}~\cite{scheibert2008}, Rubinstein and co-workers \cite{Rubinstein2006a} and Krick \textit{et al.}~\cite{Krick2012}. As discussed by Dietrich and Kilgore, the analysis of the images can be complicated by various optical scattering and resolution effects (especially at the edges of microcontacts) which requires appropriate deconvolution procedures if one wants to get a quantitative measurement of the actual contact area from contact images. Here, contact images will be analyzed under the assumption that transmitted light intensity at a given contact location is proportional to the proportion of area into 
contact. As detailed below, the validity of this assumption is supported by static indentation experiments carried out at various imposed indentation depths. In order to improve the signal to noise ratio of the camera, each static contact image at a given prescribed indentation depth is obtained by averaging $300$ images. A reference image is also obtained in the same way using a non contact configuration. When subtracted to the contact image, this reference image enforces the background of the image to be almost zero thus allowing to clearly identify the size of the circular contact region. For each pixel, a normalized transmitted light intensity $I_n$ is defined as follows
\begin{equation}
I_{n}=\frac{I_{c}-I_{r}}{I_{r}}\:,
\label{eq:pix_normalization}
\end{equation} 
where $I_{c}$ is the measured light intensity under contact conditions and $I_{r}$ is the corresponding intensity in the reference image (without contact). Here, it should be kept in mind that the transmitted light intensity measured at the length scale of a pixel ($5 \times 5\:\mu \text{m}^2$) is characteristic of a multicontact interface as a result of the self affine fractal nature of the glass surface. Normalized radial intensity profiles are subsequently obtained from an angular average of the normalized images with respect to an origin defined by the apex of the lens. Results are shown in Fig.~\ref{fig:intensity_normal}a where the profiles can be seen to be shifted to higher light intensity values when indentation depth is increased. Interestingly, it comes out that all the profiles obtained at various contact loads collapse onto a single plot (Fig.~\ref{fig:intensity_normal}b) when intensity data are normalized with respect to the average contact pressure $p_m=P/\pi a^2$ ($P$ is the applied normal 
load) and the radial coordinate is normalized with respect to contact radius $a$. If the local contact pressure $\sigma_{zz}$ is assumed to scale as $\sigma_{zz}(r/a) \propto P/\pi a^2 f(r/a)$ (where $f$ is some function of the space coordinate), this means that transmitted light intensity scales locally with the  applied contact pressure. This result is further illustrated by the linear relationship between the integrated light intensity transmitted through the rough contact area and the applied normal load (Fig.~\ref{fig:total_intensity}). It is noteworthy that a similar result was obtained by Rubinstein~\textit{et al.}~\cite{Rubinstein2006} using a different optical technique where a laser sheet is incident on a contact interface between two rough PMMA blocks at an angle far beyond the angle for total internal reflection from the PPMA/air interface. Under the assumption that the transmitted light intensity is proportional to the proportion of area into contact, the observation of such a linear 
relationship is consistent with many rough contact theories~\cite{persson2001,greenwood1966,carbone2008,campana2007} which predict that the actual contact area varies linearly with the applied load, at least in the low load range. Accordingly, we will make the assumption that the recorded light intensity at a given pixel location is proportional to the proportion of area into contact, $I_n \propto {A/A_0}$ where $A$ and $A_0$ are the actual and nominal contact areas, respectively. For the surface topography under consideration, this hypothesis is supported by the above reported indentation experiments even if it is not necessarily valid for any kind of roughness. \\
During steady state friction, a systematic change in the distribution of transmitted light within the contact is observed not only as a function of the applied indentation depth but also as a function of the imposed angular velocity. In order to quantify these changes, the following treatment is applied to the recorded contact images. For a given indentation depth and applied velocity, sequences of images such as that shown in Fig.~\ref{fig:contact_pictures}b are averaged. The resulting time-averaged picture is subsequently averaged as a function of the angular coordinate with respect to the center of rotation in order to get a radial profile. For normalization purposes, a light intensity profile is also obtained using the same averaging procedure with a sequence of images where the rotating lens is close to but not in contact with the rubber surface. An example of the resulting profiles is shown in Fig.~\ref{fig:intensity_profiles} for an indentation depth of 140~$\mu$m and various velocities ranging from 0.
01 to 1~deg~s$^{-1}$. At a given location within the contact, i.e. for a given contact pressure, it turns out that the amount of light transmitted locally through the rough contact interface is decreasing as the local sliding velocity is increased. There is thus some evidence that micro-contacts at the frictional interface are redistributed as a function of the sliding velocity, more precisely that the proportion of area in contact decreases  at high sliding velocities.
Recalling the assumption that light intensity is proportional to the proportion of area into contact, i.e. $I_n(p,v) \propto A(p,v)/A_0$ the dependence of the frictional shear stress on the actual contact area should therefore be reflected by the ratio $\tau(p,v)/I_n(p,v)$. When this ratio is plotted as a function of the local sliding velocity, it comes out that all the data point obtained at various imposed angular velocity and applied indentation depth merge on a single master curve (Fig.~\ref{fig:kv}). Remarkably, this master curve is independent on the contact pressure (i.e. on both the location within the contact and on the imposed indentation depth). From this observation, the measured local shear stress can thus be expressed in the following way
\begin{equation}
\tau(p,v)=k(v)I_n(p,v)\propto k(v)A(p,v)/A_0\:.
\label{eq:tau_k_I}
\end{equation}
\section{Discussion}
\label{sec:Discussion}
From the inversion of the displacement field at the surface of a viscoelastic rubber contacting a rigid spherical asperity, local values of the steady state frictional shear stress were determined under torsional contact conditions. For rough contacts, measured values of the local shear stress are representative of a multicontact interface under given sliding velocity and nominal contact pressure conditions. In addition, information about the local proportion of area into contact $A/A_0$ is provided from contact images under the assumption that optical transmittivity of the rough interface is proportional to $A/A_0$. As detailed above, this assumption is supported by separate static indentation measurements where it yields the 
expected linear relationship between the actual contact area and the applied nominal contact pressure. From a systematic investigation of the local shear stress and transmitted light intensity as a function of the applied indentation depth and twisting rate, it is found experimentally that the frictional shear stress can be expressed as the product of two terms (cf Eqn(\ref{eq:tau_k_I})). The first one, $A/A_0$, incorporates the velocity and pressure dependence of the microcontacts density. The second one is a pressure-independent term, $k(v)$, which can be viewed as some averaged measurement of the amount of frictional energy dissipated within micro-contacts. In other words, $A/A_0$ corresponds to a contact mechanics term describing the density of microcontacts under steady state sliding while $k(v)$ quantifies the dissipative processes at play within asperity micro-contacts.\\
As shown by the dotted line in Fig.~\ref{fig:kv}, it can interestingly be noted that the magnitude of $k(v)$ is similar to that of the frictional shear stress measured for smooth contacts and that it follows a very similar velocity dependence. This suggests that frictional energy dissipation within microasperity contacts is mostly due to interfacial dissipation, the contribution of viscoelastic losses at asperity scale being negligible. This statement can be further considered within the framework of the friction model detailed in the introduction. Accordingly, the frictional force is assumed to arise from two independent contributions, namely the so-called adhesive and hysteretic components. The so-called adhesive term encompasses all dissipative mechanisms occurring at the points of intimate contact between the solids, i.e. on length scales lower than asperity size. The hysteretic term corresponds to the force required to displace the rubber material from the front of the rigid nano-asperities. Here, it 
represents the contribution of the viscoelastic losses involved in the deformation of the rubber substrate by microasperities. Rewritten in terms of shear stress, this model can be expressed as follows
\begin{equation}
\tau=\tau_h + \tau_a\: ,
\end{equation}
where $\tau_a$ and $\tau_h$ are respectively the adhesive and hysteretic terms. The adhesive term can simply be expressed as 
\begin{equation}
\tau_a=\tau_0 \frac{A}{A_0}\;
\label{eq:taua}
\end{equation}
where $\tau_0$ is the frictional shear stress of the smooth contact interface.\\
An exact calculation of the hysteretic component $\tau_h$ is much more complicated as it implies to solve the viscoelastic contact problem taking into account the whole frequency distribution associated with the topography of the self affine rough surface. As a first order approximation, we follow a simple approach where the rough surface is assimilated to a distribution of identical, non interacting, spherical asperities. Following a calculation by Greenwood and Tabor~\cite{greenwood1958}, the friction force at the scale of a single asperity can be expressed as
\begin{equation}
F_{asp}=\alpha\frac{E_{eff}}{4}\frac{a^4}{R^2}\,
\end{equation}
where $R$ is the radius of curvature of the asperity and $E_{eff}$ is a frequency dependent effective modulus defined as
\begin{equation}
E_{eff}(\omega)=\frac{\left|E(\omega)\right|}{1-\nu^2}
\end{equation}
where $\omega$ is a characteristic frequency defined as $\omega=v/a$ and $\nu$ is the Poisson's ratio whose variations with frequency are neglected. In the above equation, $\alpha$ is term representing the fraction of the input elastic energy which is lost as a result of viscoelastic dissipation. The hysteretic frictional stress can thus be written as $\tau_h=\phi F_{asp}$ where $\phi$ denotes the surface density of asperities. $\phi=A/(\pi a^2 A_0)$ thus allowing to express $\tau_h$ as
\begin{equation}
\tau_h=\alpha \frac{E_{eff}}{4 \pi}\frac{A}{A_0}\left(\frac{a}{R}\right)^2\:.
\end{equation}
As an upper bound value for $\tau_h$, one can take $a\approx R$ which gives
\begin{equation}
\tau_h \approx \alpha \frac{E_{eff}}{4 \pi}\frac{A}{A_0}\:.
\label{eq:tauv}
\end{equation}
From eqns (\ref{eq:taua}) and (\ref{eq:tauv}), the total frictional stress within the rough interface can thus be expressed as
\begin{equation}
\tau=\frac{A}{A_0} \left( \tau_0 + \alpha \frac{E_{eff}}{4 \pi} \right)\:.
\label{eq:tautot}
\end{equation}
Within the investigated sliding velocity range (0.1~to~100~$\mu$m~s$^{-1}$), the adhesive term $\tau_0$ is found to vary between 0.2 and 0.5~MPa (Fig.~\ref{fig:master_smooth}). The estimate of the second, viscoelastic term, in the RHS of Eqn~\ref{eq:tautot} requires a knowledge of the dissipation factor $\alpha$. Following an exact viscoelastic calculation by Persson~\cite{persson2010}, we take for $\alpha$ an asymptotic (low velocity) value calculated as $\alpha \approx 5 \tan \delta$ where $\tan \delta$ is the loss tangent of the rubber substrate. Using this approach and the viscoelastic data reported in the appendix, the viscoelastic term $\alpha E_{eff}/4 \pi$ is found to vary between 0.05 and 0.1 MPa when the characteristic frequency varies between 0.1~Hz and 1~kHz. This simple calculation thus yields an estimate of the hysteretic term which is found to be about half the magnitude of the interface term. The rough approximations embedded in the calculation do not really allow to draw a definite conclusion from a difference of less than one order of magnitude. Here, it can just be stated that the above calculation does not contradict the fact that the interfacial contribution to friction could be the dominant term, as suggested by the similarity between $k(v)$ and $\tau_0(v)$. 
However, this calculation is based on a very crude description of the contact interface which is assumed to consist of a distribution of identical, non interacting, single-asperity contacts. As a result, topographical features of the surface such as rms roughness, fractal dimension or correlation length are not taken into account. A more refined approach to the hysteretic component to friction would require that the multiscale features of surface topography as well as non linear effects encountered during deformation at microasperity scale are accounted for. Some of these features are embedded within theoretical rough contact models such as that developed by Persson~\cite{persson2001} but using these models would require extensive calculations which are beyond the scope of this study. From an experimental perspective, more insights into the adhesive and hysteretic components to friction could be gained from experiments where the physical chemistry of the glass surface is varied (using silanization for example) independently of the viscoelastic properties of the rubber or, conversely, where the viscoelasticity of the substrate is changed independently of the properties of the glass surface. When doing so, one should take care to the potential occurrence of stick-slip motions or friction instabilities which would preclude such an analysis.
\section{Conclusion}
\label{sec:conclusion}
Using contact imaging approaches, we were able to determine the distribution of frictional stresses within contacts between a smooth viscoelastic rubber and a rigid rotating lens. When the lens surface is made randomly rough, the local frictional stress is observed to dependent on both contact pressure and on sliding velocity as a result of the multicontact nature of the sliding interface. Associated changes in the local density of micro-contacts are evidenced from variations in the distribution of the light intensity transmitted through the rough contact interface. From separate static indentation experiments, it is shown that all the light intensity data obtained locally at various contact loads and contact locations can be represented in the form of a single master curve which strongly supports the scaling of the transmitted light with the nominal contact pressure, at least for the considered rough surface. Accordingly, the theoretical prediction of a linear relationship between the proportion of area in contact and contact pressure is retrieved experimentally. More importantly, the combination of local stress and light intensity measurements allowed to separate the contributions of two mechanisms when contact pressure or velocity is varied. The first one consists in a decrease in the local density of micro-contacts when the pressure decreases. The second one encompasses all the frictional dissipative processes occurring within microasperity contacts. A comparison between smooth and rough contacts suggests that dissipative processes occurring at the interface predominate over viscoelastic dissipation at micro-asperity scale. More generally, these results open the way to a close reexamination of the validity of the hypothesis embedded in most rubber friction models, especially the assumption that friction can be separated into an adhesive and an hysteretic component. %
\begin{acknowledgments}
	This work was partially supported by the National Research Agency (ANR) within the framework of the Dynalo project (NT09499845). The authors wish to thank Basile Pottier and Laurence Talini for the surface fluctuations measurements. Thanks are also due to Danh Toan Nguyen for the finite element calculations reported in the appendix. We are also indebted to Alexis Prevost for many stimulating discussions.\\
\end{acknowledgments}
\appendix
\section{Linear viscoelastic measurements}
\label{app:visco_measures}
\begin{figure}
	\includegraphics[width=\columnwidth]{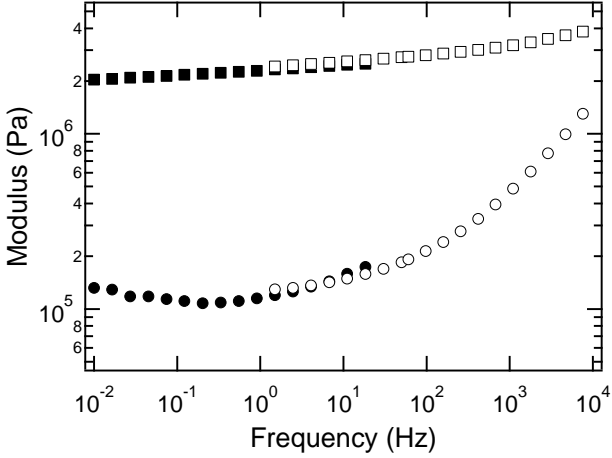}
	\caption{Frequency dependence of the room temperature storage ($G'$, squares) and loss ($G''$, circles) shear moduli of the epoxy rubber as determined from conventional DMTA (filled symbols) and SFSR spectroscopy (open symbols)}.
	\label{fig:rheoed2003}
\end{figure}

The selected epoxy rubber is characterized by a crystallization of the flexible chains of the polyether-diamine crosslinker at low temperature (-$20\,^{\circ}\mathrm{C}$). As a result, it is not possible to determine the room temperature viscoelastic modulus of the rubber over an extended frequency range using the usual route of master curves and time-temperature superposition principle. Instead, we used two complementary techniques to determine the frequency dependence of the viscoelastic modulus at room temperature. Up to 20 Hz, the shear modulus was measured using conventional Dynamical Mechanical Thermal Analysis (DMTA). Elastomer disks 2 mm in thickness and 8 mm in diameter are sheared at low strain (0.05 \%) between the parallel plates of a rheometer (Anton Paar, MCR 501). The shear modulus is measured at room temperature during a frequency sweep between 50 and 0.01 Hz. In the high frequency range (up to 10 kHz), the viscoelastic modulus is measured using Surface Fluctuation Specular Reflection (SFSR) 
spectroscopy, a technique based on the principle that surface fluctuations reveal the properties of the medium. The principle of this technique is fully described in references~\cite{pottier2013,pottier2011}. The results of both viscoelastic measurements are shown in Fig.~\ref{fig:rheoed2003}. %
\section{Inversion of the displacement field: comparison between Green's tensor and Finite Element calculations}
\begin{figure}
	\includegraphics[width=\columnwidth]{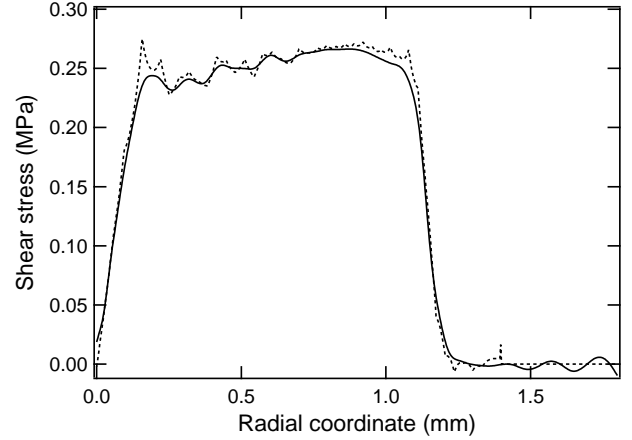}
	\caption{Shear stress derived from the inversion of the same displacement field using either a linear contact mechanics approach (continuous line) or a FE calculation taking into account the geometrical and material non linearities (dotted line).}
	\label{fig:green_FEM}
\end{figure}

In order to evaluate the influence of finite strains on the inversion of displacement fields, the same azimuthal displacement profile was inverted using both a linear elastic approach based on Green's tensor~\cite{chateauminois2008} and a Finite Element (FE) inversion procedure which is fully described in reference~\cite{nguyen2011}. As opposed to Green's tensor calculations, the FE inversion is able to take into account both the geometrical and material non linearities (neo-Hookean behavior of the rubber) of the problem. As shown in Fig.~\ref{fig:green_FEM}, identical shear stress profiles are provided by both methods. In other words, the occurrence of finite strains at the edge of the contact (see Fig.~\ref{fig:displ_field}) does not induce any significant error in the stress field deduced from an inversion using a linear elastic analysis. It should be noted that this conclusion is opposed to that drawn for linear sliding conditions: in this case, finite strains were found to alter significantly the accuracy of linear inversions~\cite{nguyen2011}. Some theoretical justifications for this difference can be found in finite strain analytical calculations~\cite{huy2013}.
%
\bibliographystyle{unsrt}

%
%
%
\end{document}